\documentclass[epj]{webofc}
\usepackage[utf8]{inputenc}
\usepackage[varg]{txfonts} % Web of Conferences font
\usepackage{booktabs}
\usepackage{xcolor}
\definecolor{darkred}{rgb}{0.4,0.0,0.0}
\definecolor{darkgreen}{rgb}{0.0,0.4,0.0}
\definecolor{darkblue}{rgb}{0.0,0.0,0.4}
\usepackage[bookmarks,linktocpage,colorlinks,
linkcolor = darkred,
urlcolor= darkblue,
citecolor = darkgreen]{hyperref}
\usepackage{subcaption}
\usepackage{slashed} % f"ur slash-Notation
\usepackage{bbm} % f"ur Identit"at \mathbbm{1}
\usepackage{siunitx}
\usepackage{units}
\usepackage[autostyle=true]{csquotes}
\usepackage{mathrsfs}
\wocname{EPJ Web of Conferences}
\woctitle{Lattice2017}
\newcommand{\ii}{\mathrm{i}}
\newcommand{\e}{\mathrm{e}}
\newcommand{\tr}{\mathrm{tr}}
\begin{document}
%%%%%%%%%%%%%%%%%%%%%%%%%%%%%%%%%%%%%%%%%%%%%%%%%%%%%%%%%%%%%%%%%%%%%%%%%%%%%
%
\selectlanguage{english}
%----------------------------------------------------------------------------
\title{%
Spectroscopy of four-dimensional $\mathcal{N}=1$ supersymmetric SU(3) Yang-Mills theory
}
%----------------------------------------------------------------------------
\author{%
\firstname{Marc} \lastname{Steinhauser}\inst{1}\fnsep\thanks{Speaker, \email{marc.steinhauser@uni-jena.de}} \and
\firstname{Andr\'e} \lastname{Sternbeck}\inst{1} \and
\firstname{Bj\"orn} \lastname{Wellegehausen}\inst{1,2} \and
\firstname{Andreas} \lastname{Wipf}\inst{1}
}
% TODO all e-mails?
%----------------------------------------------------------------------------
\institute{%
Theoretisch-Physikalisches Institut, Friedrich-Schiller-Universit\"at Jena, 07743 Jena, Germany
\and
Institut für Theoretische Physik, Justus-Liebig-Universit\"at Gie{\ss}en, 35392 Gie{\ss}en, Germany
}
%----------------------------------------------------------------------------
\abstract{%
	Supersymmetric gauge theories are an important building block for extensions of the standard model. As a first step towards Super-QCD we investigate the pure gauge sector 
	with gluons and gluinos on the lattice, in particular the low energy mass spectrum: meson-like gluinoballs, gluino-glueballs and pure glueballs. We report on some first calculations performed with clover improved Wilson fermions on rather small lattices. The supersymmetric continuum limit and particle masses are discussed and compared to predictions from effective field theory.
}
%----------------------------------------------------------------------------
\maketitle
%----------------------------------------------------------------------------
\section{Introduction}\label{ch:intro}
The standard model of particle physics successfully 
describes and predicts observations at low energies. Nevertheless 
it is only an effective theory with several weak points like the
hierarchy problem and the infinite vacuum energy. These and further
deficiencies motivate 
the search for a more fundamental theory of particle physics. 
One possibility is the extension of the standard model with supersymmetry. Here we focus on the strong interaction sector and study an important building block of supersymmetric quantum chromodynamics (Super-QCD), the four-dimensional $\mathcal{N}\!=\!1$ Super-Yang-Mills theory (SYM) with gauge group SU(3). Previous lattice studies on this model were performed by the DESY-M\"unster collaboration, see~\cite{Bergner1512} for a summary of their results on SYM with gauge group SU(2) and \cite{Ali1610,Ali1710} for first investigations with gauge group SU(3).

The $\mathcal{N}\!=\!1$ SYM theory contains only two fields: 
a gauge potential $A_\mu(x)$ and a Majorana field $\lambda(x)$,
both transforming with the adjoint representation. They
describe a gauge boson, called gluon and its superpartner, 
named gluino. The gluon and gluino are members of 
the same super-multiplet and hence
have the same number of propagating degrees of freedom.
The supersymmetry transformation relating the fermion field
and boson field reads
\begin{equation}
	\delta_\epsilon A_\mu = \ii \bar\epsilon\gamma_\mu\lambda,~~~\delta_\epsilon \lambda= \ii \Sigma_{\mu\nu}F^{\mu\nu}\epsilon \nonumber\,,
\end{equation}
where $F_{\mu\nu}$ is the field strength tensor, $\Sigma$ is defined as $\Sigma_{\mu\nu}\equiv\frac{\ii}{4}[\gamma_\mu,\gamma_\nu]$ and $\epsilon$ is an infinitesimal anti-commuting constant Majorana spinor. 
The continuum on-shell Lagrange density can be written as
\begin{equation}
	\mathcal{L}_\text{SYM}=\tr\left( -\frac{1}{4}F_{\mu\nu}F^{\mu\nu}+\frac{\ii}{2}\bar{\lambda} \slashed{D} \lambda -\frac{m}{2}\bar{\lambda}\lambda \right)\,.
\end{equation}
This formulation contains a gluino mass term which breaks
supersymmetry softly. We will use this term in the Wilson lattice
formulation to fine-tune our theory to the physical point at
vanishing gluino mass.

Already in the 80s, the particle spectrum of $\mathcal{N}\!=\!1$ SYM theory was studied within the framework of effective field theory using the theory's symmetries and the anomaly matching. At
high energies, gluons and gluinos can be treated as a free gas, like
in Yang-Mills theory. Furthermore, we encounter confinement at low
energies in both theories. As a result, no free colour charges can
be observed and instead only colour-neutral states are measurable.
There are three different types of bound states: the pure
glueballs, the pure gluinoballs and the gluino-glueballs consisting
of gluons and gluinos.

Based on supersymmetry, theses bound states will be arranged in supermultiplets. Veneziano and Yankielowicz predicted in~\cite{Veneziano8206} a first multiplet:
\begin{table}[h]
	\centering
	\begin{tabular}{lll} \toprule
		1 bosonic scalar & $0^{++}$ gluinoball & $\text{a-}f_0\sim\bar{\lambda}\lambda$ \\
		1 bosonic pseudoscalar & $0^{-+}$ gluinoball & $\text{a-}\eta^\prime\sim\bar{\lambda}\gamma_5\lambda$ \\
		1 majorana-type spin $\frac{1}{2}$ & gluino-glueball & $~~\chi~~\sim F_{\mu\nu}\Sigma^{\mu\nu}\lambda$ \\\bottomrule	
	\end{tabular}
\end{table}
\newline
The nomenclature is in analogy to QCD, with the prefix \enquote{a-} indicating the adjoint representation and with the quantum numbers $J^{pc}$, for the total angular momentum $J$, the parity $\text{p}=\pm1$ and the charge conjugation $\text{c}=\pm1$. Later, Farrar, Gabadadze and Schwetz introduced a further multiplet by including an additional term in the action to maintain the states corresponding to glueballs~\cite{Farrar9711}. This results in a second supermultiplet which contains:
\begin{table}[h]
	\centering
	\begin{tabular}{lll} \toprule
		1 bosonic scalar & $0^{++}$ glueball & $0^{++}\sim F_{\mu\nu}F^{\mu\nu}$ \\
		1 bosonic pseudoscalar & $0^{-+}$ glueball & $0^{-+}\sim \epsilon_{\mu\nu\rho\sigma}F^{\mu\nu}F^{\rho\sigma}$ \\
		1 majorana-type spin $\frac{1}{2}$ & gluino-glueball & $\,~\chi~~\sim F_{\mu\nu}\Sigma^{\mu\nu}\lambda$ \\\bottomrule	
	\end{tabular}
\end{table}

In comparison to QCD, the SYM theory possesses a different breaking pattern of chiral symmetry. When the gluino mass term is absent the SU($N_\text{c}$) SYM theory admits a global chiral U$(1)_\text{A}$ symmetry, $\lambda \mapsto \e^{\ii\alpha\gamma_5}\lambda$.
Due to the anomaly, however, only a $\mathbb{Z}_{2 N_\text{c}}$ remnant symmetry
\begin{equation}
	\lambda \mapsto \e^{ 2\pi\ii n\gamma_5/2 N_\text{c}}\lambda ~~~\text{with}~~~ n\in\{1,\ldots,2 N_\text{c} \}
	\nonumber
\end{equation}
survives. As a consequence of a gluino condensate $\langle \bar{\lambda}\lambda \rangle\neq 0$, this remnant symmetry is 
further broken spontaneously to a $\mathbb{Z}_2$ symmetry. 
Therefore the theory contains $N_\text{c}$ physically equivalent
vacua.

To investigate the mass spectrum and confinement, non-perturbative methods like lattice simulations are required. In our work, the method of Curci and Veneziano is applied, where supersymmetry is broken explicitly on the lattice by using Wilson fermions~\cite{Curci8612}. The symmetry breaking results in a counter term proportional to the gluino mass term. To solve this
problem, an explicit gluino mass term is added and its bare mass is 
fine-tuned such that the gluino becomes massless in the continuum limit. Due to confinement, the gluino can't be measured directly to find the correct parameter. As proposed in \cite{Veneziano8206}, we circumvent this problem by measuring the adjoint pion mass squared,
$m_{\text{a-}\pi}^2\propto m$. This observable needs only few statistics and is computationally cheap to determine. In the SYM theory, the adjoint pion isn't a physical particle; it is defined in a partially quenched approximation similarly to 1-flavour QCD. With fine-tuning to the critical gluino mass $m_\text{crit}$ we can assure at the same time the restoration of supersymmetry, as well as the restoration of chiral symmetry, in the continuum limit.

In comparison to QCD, we encounter in the path integral the Pfaffian
of the Dirac operator after integrating out the Majorana fermions.
Thus, we perform our simulations with the rational hybrid Monte Carlo
algorithm (RHMC)~\cite{Kennedy9809}.

\section{Clover operator}\label{ch:artefacts}
In order to arrive at results for the continuum theory, multiple
issues must be considered in lattice calculations. Simulations have
to be performed near the critical point, where the adjoint pion
and gluino have minimal (lattice) mass. The chiral limit is 
necessary to restore chiral symmetry as well as supersymmetry in the
continuum limit. Then, the thermodynamic limit is performed to eliminate finite volume effects. Finally, the continuum limit is the
last step to restore the continuum symmetries.

To reduce the discretization errors of our simulation, we use improved lattice actions~\cite{Symanzik83I,Symanzik83II}. For the gauge part we employ the Symanzik improved L\"uscher-Weisz gauge action
\vspace{2mm}
\begin{equation}
	S_\text{g}[\mathcal{U}]=\frac{\beta}{N_\text{c}}\bigg( \frac{5}{3}\sum_{\square}\tr(\mathbbm{1}-\text{Re}\,\mathcal{U}_\square) - \frac{1}{12}\sum_{\square\square}\tr(\mathbbm{1}-\text{Re}\,\mathcal{U}_{\square\square}) \bigg)
\end{equation}
with $\mathcal{O}(a^2)$ lattice artefacts. Fermions are simulated with the Wilson-Dirac operator
\vspace{2mm}
\begin{equation}
	D_\text{W}(x,y)=\delta_{x,y} - \kappa\sum_{\mu=\pm 1}^{\pm 4}\left( \mathbbm{1}-\gamma_\mu \right) \mathcal{V}_\mu(x)\,\delta_{x+\mu,y}\,,
\end{equation}
which contains the abbreviation $\gamma_{-\mu}\equiv-\gamma_\mu$ and the gauge potential in the adjoint representation 
$\left[ \mathcal{V}_\mu(x)\right] _{ab}=2\,\tr\left[ \mathcal{U}_\mu^\dagger(x)T_a \mathcal{U}_\mu(x) T_b \right]$ calculated with the fundamental generators $T_a$. 
This fermion action has $\mathcal{O}(a)$ lattice artefacts, which can be reduced by introducing the irrelevant clover term
\vspace{2mm}
\begin{equation}
	D_\text{W}(x,y)=\delta_{x,y} - \kappa\sum_{\mu=\pm 1}^{\pm 4}\left( \mathbbm{1}-\gamma_\mu \right) \mathcal{V}_\mu(x)\,\delta_{x+\mu,y}-c_\text{SW}\frac{\kappa}{2}\Sigma_{\mu\nu}F^{\mu\nu}\delta_{x,y}\,,
\end{equation}
where $F^{\mu\nu}$ is the clover-shaped sum of 4 plaquettes. When the Sheikholeslami-Wohlert coefficient $c_\text{SW}$ is chosen properly the lattice artefacts are reduced to $\mathcal{O}(a^2)$~\cite{Luscher84,Sheikholeslami85,Wohlert87}.

The coefficient $c_\text{SW}$ is commonly determined using lattice perturbation theory~\cite{Sheikholeslami85}. Within this framework, quantities like quark-gluon vertices for massless fermions are investigated and the clover coefficient is chosen such that the $\mathcal{O}(a)$ discretization errors vanish at a certain loop order, typically at 1-loop. This method depends on the fermion representation as well as the gauge action. A general one-loop perturbative result for the Wilson gauge action and fermions in an arbitrary representation 
was derived in~~\cite{Musberg1304}:
\vspace{2mm}
\begin{equation}
	c_\text{SW}=1+g^2\,\bigg( \num{0.16764(3)}\,C_\text{rep} + \num{0.01503(3)}\, N_\text{c}\bigg)\nonumber\,.
\end{equation}
For the fundamental representation the quadratic Casimir invariant
$C_\text{rep}$ is $C_\text{fund}=(N_c^2-1)/2N_\text{c}$
and for the adjoint representation respectively
$C_{\text{adj}}=N_\text{c}$.

Tadpole improvement provides a further opportunity to adjust the clover coefficient~\cite{Lepage9209}. The mean link can be defined gauge-invariantly as $u_0\equiv\big\langle \tr ( U_\text{plaq})/N_\text{c}\big\rangle^{\nicefrac{1}{4}}$. 
A short calculation leads to the tree-level approximation $c_\text{SW}=u_0^{-3}$. In the plots below, this choice corresponds to values \mbox{$c_\text{SW}\in[2.09,2.16]$}, which are highlighted in the overview plots with a green line.

For a nonperturbative determination of $c_\text{SW}$, the Schr\"odinger functional can be employed~\cite{Luscher9609}. In this procedure, the partially conserved axial current relation (PCAC) is probed to find the coefficient $c_\text{SW}$. On the lattice, there exists an additional contribution of order $\mathcal{O}(a)$, which has to vanish for the correct clover coefficient.

Here, we apply an heuristic approach to find the value of the clover
coefficient. This idea is based on a distinctive feature of the SYM
theory. The lattice mass of the adjoint pion 
$m_{\text{a-}\pi}$ is not physical and has
to vanish in the continuum limit. Consequently, it is only an effect
of broken supersymmetry and chiral symmetry. Thus we are looking for
the parameter $c_\text{SW}$, where $m_{\text{a-}\pi}$ is minimal,
to find the point which is closest to the continuum with respect to
discretization errors and symmetries.

\begin{figure}
	\begin{subfigure}[h]{0.4\textwidth}
		\vspace{8mm}
		\hspace{0.3mm}\resizebox{1.05\textwidth}{!}{\input{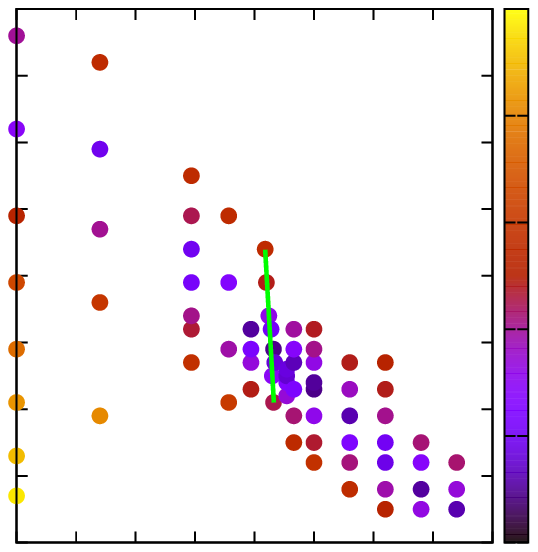}}
	\end{subfigure}
	\begin{subfigure}[h]{0.6\textwidth}
		\hspace{0.07\textwidth}
		\resizebox{0.91\textwidth}{!}{\input{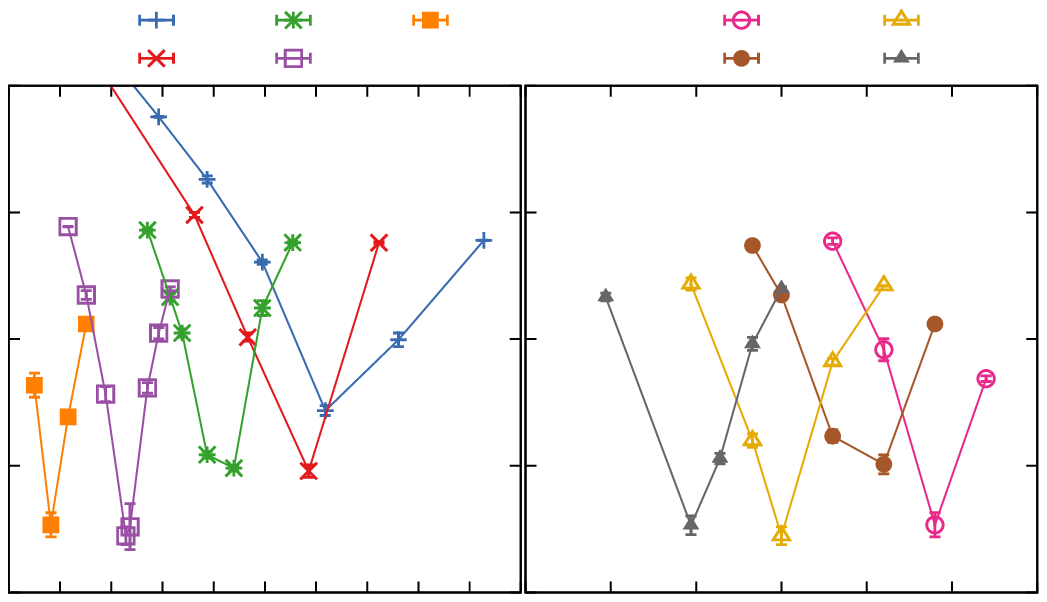}}
	\end{subfigure}
	\begin{subfigure}[h]{0.4\textwidth}
		\centering
		\resizebox{1.05\textwidth}{!}{\input{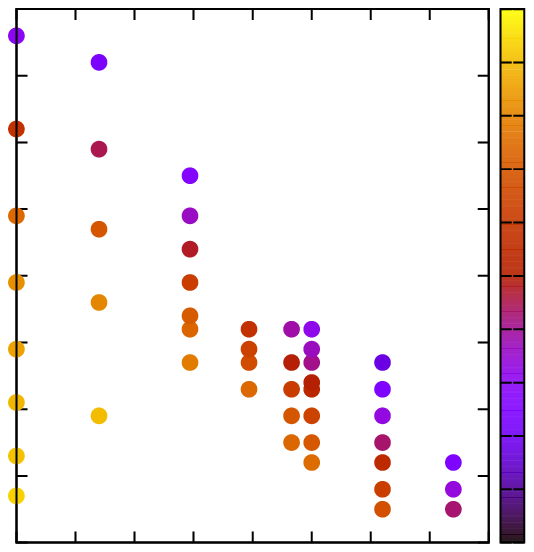}}
	\end{subfigure}
	\begin{subfigure}[h]{0.6\textwidth}
		\hspace{0.07\textwidth}
		\resizebox{0.91\textwidth}{!}{\input{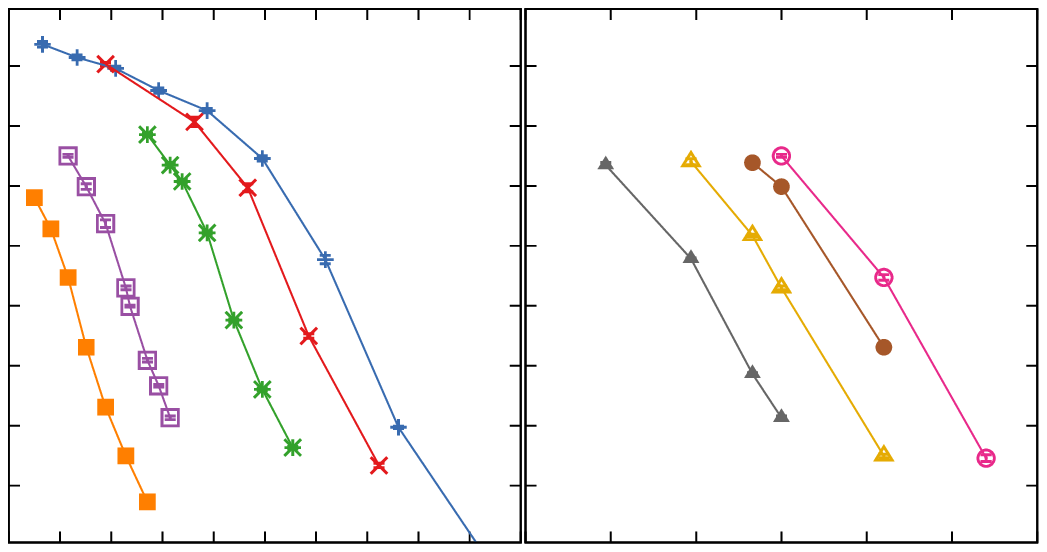}}
	\end{subfigure}
	\begin{subfigure}[h]{0.4\textwidth}
		\centering
		\resizebox{1.05\textwidth}{!}{\input{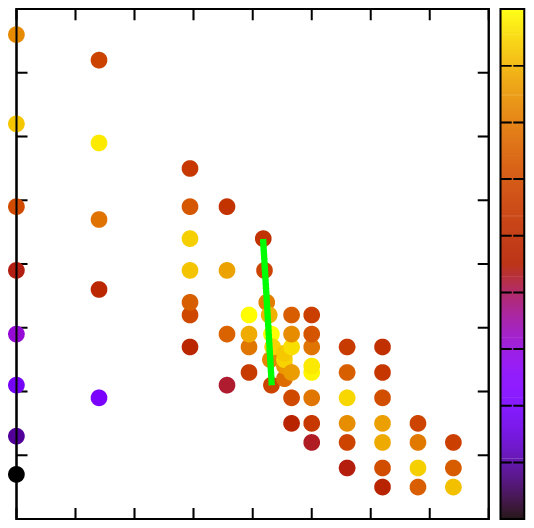}}
	\end{subfigure}
	\begin{subfigure}[h]{0.6\textwidth}
		\hspace{0.07\textwidth}
		\resizebox{0.91\textwidth}{!}{\input{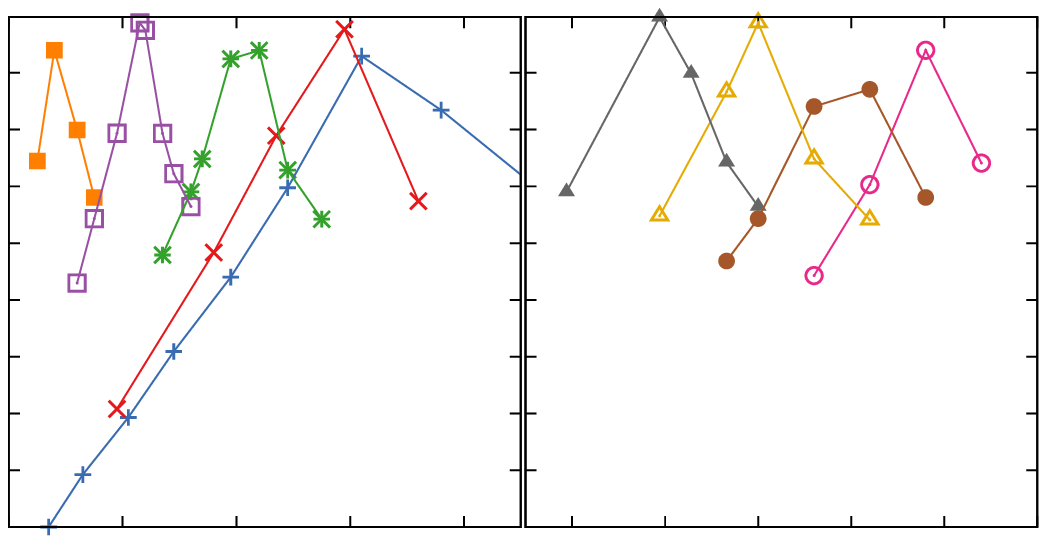}}
	\end{subfigure}
	\caption{Parameter scan for clover fermions. The top row shows the adjoint pion mass $m_{\text{a-}\pi}$. The left plot shows an overview, where the adjoint pion mass is encoded in the colour. The green line in the left plot corresponds to simulations with tadpole improved clover coefficients. The middle and right plot present the same data as cuts through \mbox{$c_\text{SW}=\text{const.}$} respectively \mbox{$m=\text{const}$}. The middle row shows the bosonic action $S_\text{B}$ and the bottom row depicts the subtracted chiral condensate \mbox{$\chi^\text{sub}$ (cf. eq.\,\eqref{eq:chiralCond})} in the same manner. Lines are included to guide the eye.}
	\label{fig:variousObserables}
\end{figure}

We vary the coefficient $c_\text{SW}\in[0,3.7]$ and chose in each case multiple hopping parameters around the critical point, where the adjoint pion mass is as small as possible. All simulations in this section were carried out on a $8^3\times 16$ lattice and at a fixed lattice coupling $\beta=4.2$. The parameter scan is shown in the top left plot of Fig.~\ref{fig:variousObserables}, where the adjoint pion mass is encoded in the colour scale. The top middle plot of this figure presents the same data for five selected $c_\text{SW}$ values. These $\kappa$-scans indicate a shift of the critical point towards lower $\kappa$ for raising $c_\text{SW}$. In the right panel, the data is represented as $c_\text{SW}$-scans for four different hopping parameters $\kappa$. We conclude that all values of $c_\text{SW}$ allow to reach the respective critical point with comparable adjoint pion mass. Altogether, there are at least three interesting choices for a lattice simulation:
\begin{itemize}
	\item without clover improvement and variable hopping parameter,
	\item with fixed chiral tree-level hopping parameter $\kappa=0.125$ and variable $c_\text{SW}$ or 
	\item between these extreme choices at the tadpole value $c_\text{SW}\approx 2.1$ with variable $\kappa$.
\end{itemize}
So far, the adjoint pion does not provide enough evidence for any 
particular choice and we need further observables. To probe a
sufficiently large parameter range, we focus on observables which 
need only a small
number of configurations. As bosonic observable we use the bosonic part of the action.\footnote{The bosonic action is part of the Ward identity $\langle S_\text{B}\rangle = \frac{3}{2} N_\text{G} V$, where $N_\text{G}$ is the number of generators of the gauge group~\cite{Catterall0811,August1710}. For the adjoint representation of SU$(3)$, this results in $\langle S_\text{B}\rangle/V = 12$, used for comparison with the broken supersymmetry at finite lattice spacing.}
Furthermore we use fermionic observables like the chiral condensate and the adjoint $a$ meson mass (its correlator corresponds to the connected part of the correlator of $\text{a-}f_0$).

Fig.~\ref{fig:variousObserables} shows two further observables in the same
manner as before. The  bosonic part of the action
(middle) is approximately constant along the diagonal running critical
line. Perpendicular to this direction, the bosonic action is decreasing with increasing values of $c_\text{SW}$
and $\kappa$. A similar behaviour can be seen for the subtracted chiral
condensate (bottom),
\vspace{2mm}
\begin{equation}
\chi^\text{sub}\equiv\chi^\text{}-\frac{m_{\text{a-}\pi}^\text{}}{m_{\text{a-}\pi}^\text{ref}}\,\chi^\text{ref}\,,
\label{eq:chiralCond}
\end{equation}
and the adjoint $a$ mass $m_{\text{a-}a}$, which is almost identically to the adjoint pion (top row of Fig.~\ref{fig:variousObserables}). These quantities are extremal along the critical line and in-/decrease in both perpendicular directions.

Unfortunately, no conclusive result can be extracted from our observables
yet. To this end, we have to vary the lattice spacing and possibly also study further observables. A final answer could give the PCAC relation,
which is sensitive to discretization artefacts. This is left for a future study.

\section{Particle masses}\label{ch:masses}
\begin{figure}[h]
	\hspace{0.06\textwidth}
	\resizebox{0.93\textwidth}{!}{\input{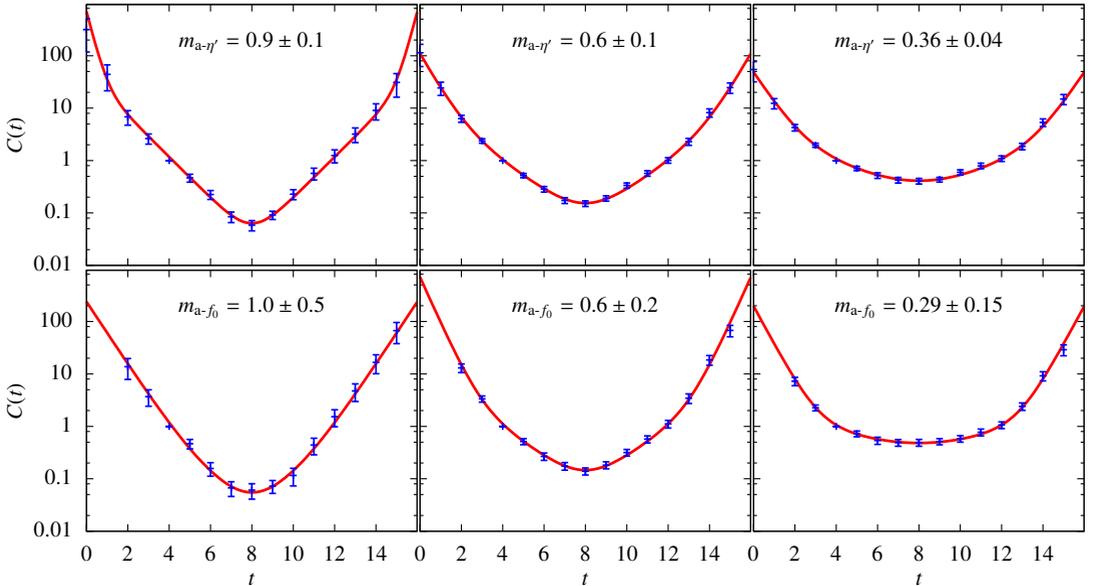}}
	\vspace{6mm}
	\caption{Correlator for adjoint meson states $\text{a-}\eta^\prime$ and $\text{a-}f_0$ on approximatively 1500 configurations with fixed improved clover coefficient $c_\text{SW}=2.27$ on the $8^3\times 16$ lattice at lattice coupling $\beta=5.2$. From left to right, the hopping parameter has the values \mbox{$\kappa=\{0.142,0.144,0.45\}$} approaching the critical point. The corresponding masses of the adjoint pion are \mbox{$m_{\text{a-}\pi}=\{0.9\pm0.1,0.6\pm0.1,0.36\pm0.08\}$}.}
	\label{fig:results--clovervalue=2.27}
	\vspace{-2mm}
\end{figure}

In this section, we present first results for the low energy mass
spectrum. Our attention is focused on the meson-type states, namely the
adjoint $\eta^\prime$ and the adjoint $f_0$. Some lattice results from simulations on a
$8^3\times 16$ lattice with an ensemble size of approximately 1500 configurations are
shown in Fig.~\ref{fig:results--clovervalue=2.27}. To
obtain meaningful results, we need to perform simulations on larger
lattices to reduce both the finite size effects and the lattice mass of
the adjoint pion, since the continuum limit is reasonable only in this
regime. Furthermore we need higher statistics to extract reliable masses,
especially for the gluino-glue states and the pure gluonic states.

\section{Outlook}\label{ch:twisted}
Recently we initiated another approach in order to come closer to
supersymmetry and chiral symmetry on comparable lattices.
In the following, we present the basic idea and preliminary 
numerical data.

The SYM theory inherits a remnant $\mathbb{Z}_{2N_\text{c}}$ symmetry
from the anomalous chiral U$(1)_\text{A}$ symmetry. Therefore, particular
directions are favored by the gluino condensate. An interesting
possibility to get closer to chiral symmetry and supersymmetry at finite
lattice spacing is given by using a parity-breaking mass resembling a twisted mass.
The mass term of the Wilson-Dirac operator breaks the
chiral symmetry and generates a condensate $\langle\bar{\lambda}\lambda\rangle$ whereas the $\mu$-mass leads to a condensate $\langle\bar{\lambda}\gamma_5\lambda\rangle$, connected to the previous by a U(1)-symmetry. A breaking in this direction can be
achieved by using a maximal twist, i.\,e., fixing the bare gluino mass at
the critical point and varying the twisted mass $\mu$ there. The
Dirac operator with twisted mass is defined as
\begin{equation}
	D_\text{W}^\text{tw}(x,y) = (4+m+\ii\,\mu\,\gamma_5)\,\delta_{x,y} - \frac{1}{2}\sum_{\mu=\pm 1}^{\pm 4}\left( \mathbbm{1}-\gamma_\mu \right) \mathcal{V}_\mu(x)\,\delta_{x+\mu,y}\,.
\end{equation}
Note that the SYM theory consists only of one flavour and consequently
the Dirac operator does not contain a $\tau_3$ matrix as in 2-flavour QCD.
With the definition of the polar mass \mbox{$M\equiv\sqrt{m^2+\mu^2}$} and the twist angle \mbox{$\alpha\equiv\arctan(\mu/m)$}, we can rewrite the mass as $m+\ii\,\mu\,\gamma_5=M\,\e^{\,\ii\,\alpha\,\gamma_5}$.

\begin{figure}[h]
	\begin{subfigure}[h]{0.62\textwidth}
		\hspace{5mm}
		\resizebox{\textwidth}{!}{\input{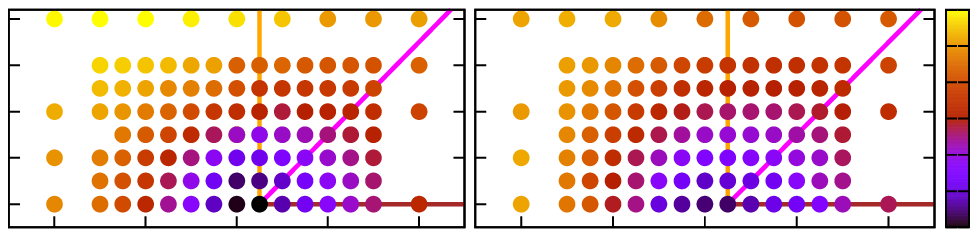}}
	\end{subfigure}	
	\begin{subfigure}[h]{0.37\textwidth}
		\hspace{2mm}
		\resizebox{\textwidth}{!}{\input{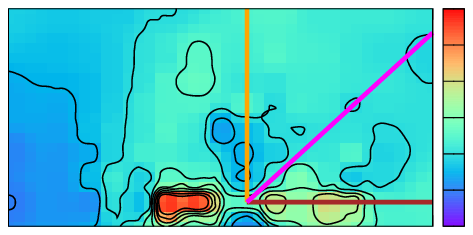}}
	\end{subfigure}	
	\begin{subfigure}[h]{\textwidth}
		\hspace{5mm}
		\resizebox{0.95\textwidth}{!}{\input{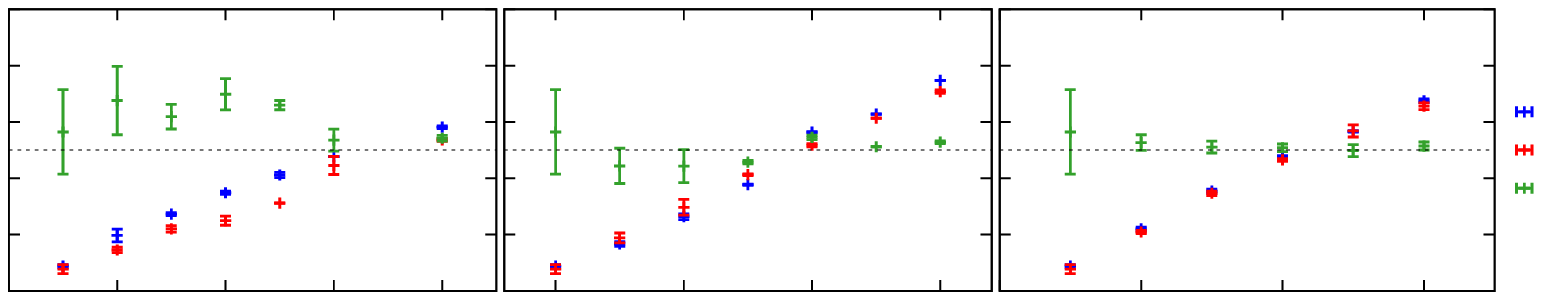}}
	\end{subfigure}
	\caption{Parameter scan for twisted mass fermions with sets of 200 configurations. From left to right, the plots in the top row show the adjoint pion mass, the adjoint $a$ mass and the ratio $m_{\text{a-}a}/m_{\text{a-}\pi}$. In the bottom row, there are slices through these plots along the horizontal, vertical \& diagonal line towards the critical point at $m_\text{crit}\approx-0.95$. Note, the observables are not rotated with the twist angle $\alpha$.}
	\label{fig:TMvariousObserables}
	\vspace{-4mm}
\end{figure}
We performed a parameter study around the critical point with \mbox{$m\in[-1.4,-0.6]$}, \mbox{$\mu\in[-0.4,0.4]$} and fixed lattice coupling $\beta=5.4$ on a $8^3\times 16$ lattice. Due to the symmetry $\mu\leftrightarrow -\mu$, detailed measurements were only performed in the upper half-plane of the parameter space.

Interesting observables are the (unphysical) $\text{a-}\pi$ and $\text{a-}a$ mesons, which correspond to the connected parts of the physical $\text{a-}\eta^\prime$ and $\text{a-}f_0$ mesons. For this first investigations we use unmodified observables, i.\,e., they are not rotated with the twist angle $\alpha$. The results are shown in Fig.~\ref{fig:TMvariousObserables} (top left \& top middle, bottom row in blue \& red) and they look almost the same, however the top right plot reveals an interesting fact.

On the one hand, along the horizontal line $\mu=0$ (brown curve) corresponding to the
untwisted scenario, the $\text{a-}\pi$ mass grows faster than the
$\text{a-}a$ mass when moving away from the critical point at
\mbox{$m_\text{crit}\approx-0.95$} (bottom left). On the other hand, $m_{\text{a-}a}$ is heavier than
$m_{\text{a-}\pi}$ along the vertical $\mu$-direction in the vicinity of the critical point (orange curve, bottom middle). A promising
way to approach the critical point is along the diagonal in the
$(m,\mu)$-space (magenta curve).
In this direction neither the $\text{a-}\pi$ 
nor the $\text{a-}a$ is preferred (bottom right). Our observations
show with good agreement that $m_{\text{a-}\pi}\approx m_{\text{a-}a}$ holds along this line. To this
effect, an improvement of the supersymmetry at finite lattice spacing
may be possible. Furthermore, the errors of the masses are rather small along this line, such that an extrapolation in this direction may be advantageous.

Overall, open questions remain in this first investigation, which may lead to interesting possibilities. There are two different ways to interpret the twisted mass. First, it can by seen as a transformation of the basis keeping the physical theory unchanged. This results in a mixing of the parity partners based on the quantum numbers:
\vspace{-3mm}
\begin{align*}
	\text{a-}a_\text{tm}&=\cos(\alpha) \cdot \text{a-}a + \ii\,\sin(\alpha)\cdot\text{a-}\eta^\prime\,, \qquad\qquad\text{a-}\pi_\text{tw}=\cos(\alpha)\cdot \text{a-}\pi + \ii\,\sin(\alpha)\cdot\text{a-}f_0\,,\\
	\text{a-}f_0^\text{tw}&=\cos(\alpha)\cdot \text{a-}f_0 + \ii\,\sin(\alpha)\cdot \text{a-}\pi\,, \qquad\qquad\text{a-}\eta^\prime_\text{tw}=\cos(\alpha)\cdot \text{a-}\eta^\prime + \ii\,\sin(\alpha)\cdot \text{a-}a_0\,,
\end{align*}
and thus observables must be modified to arrive at the physical result. Alternatively, this additional mass term may deform the physical lattice theory for a fixed basis. Nevertheless, the continuum limit of vanishing gluino mass ($m\rightarrow0,\,\mu\rightarrow0$) should lead to the same continuum theory.

So far, by observing the mesonic states we probed the chiral symmetry breaking. By investigating the gluino-glueball, we expect a deeper understanding which direction in the $(m,\mu)$-plane should be favoured in the context of supersymmetry.
Further details may be extracted from 1-loop lattice perturbation theory. In this framework, we can study if $\mathcal{O}(a)$ improvement is achievable and how the supersymmetry is broken at finite lattice spacing.

\section{Summary}\label{ch:summary}
We have studied the $\mathcal{N}=1$ supersymmetric Yang-Mills theory 
on the lattice
and included a clover term to reduce $\mathcal{O}(a)$ discretization
effects. To adjust the Sheikholeslami-Wohlert coefficient $c_\text{SW}$ we have followed an heuristic approach exploiting a special property of the theory: The
(unphysical) adjoint pion mass has to vanish in the
continuum limit. This means that this mass is purely an effect of
broken supersymmetry and chiral symmetry. Consequently, we have varied $\kappa$ and $c_\text{SW}$ to find minimal pion masses assuming at these parameters we are as close as
possible to the continuum, both with respect to discretization errors and
symmetries. For our investigations, we have used mostly a $8^3\times 16$
lattice and different observables like the
bosonic action, chiral condensate, $\text{a-}\pi$ mass and $\text{a-}a$
mass.

\subsection*{Acknowledgments}
\begin{acknowledgement}\noindent
M.\,S., A.\,S. and B.\,W. were supported by the DFG Research Training Group 1523/2 "Quantum and Gravitational Fields". B.W. was supported by the Helmholtz International Center for FAIR within
the LOEWE initiative of the State of Hesse. We gratefully acknowledge the computing time granted by the Leibniz Supercomputing Center of the Bavarian Academy of Sciences and Humanities (LRZ) on the supercomputer SuperMUC (project pr48ji). We thank Georg Bergner, Stefano Piemonte and especially Gernot M{\"u}nster for insightful discussions.
%Further computing time has been provided by the cluster of the Theoretical Physical Institute of the Friedrich Schiller University Jena.
\end{acknowledgement}

% BibTeX or Biber users please use (the style is already called in the class, ensure that the "woc.bst" style is in your local directory)
% \bibliography{name or your bibliography database}
%
\bibliography{references}

%%%%%%%%%%%%%%%%%%%%%%%%%%%%%%%%%%%%%%%%%%%%%%%%%%%%%%%%%%%%%%%%%%%%%%%%%%%%%
\end{document}